\documentclass[conference,a4paper]{IEEEtran}
\IEEEoverridecommandlockouts
\ifCLASSINFOpdf
\else
\fi

\usepackage{url}
\usepackage{graphicx,epsfig}% Include figure files
\usepackage{subfig}
\usepackage{graphics,dcolumn,bm,epic, eepic,fleqn,float}
\usepackage{amssymb,amsmath,multirow,rotate,color,mathrsfs}
\usepackage{epstopdf}
\usepackage{soul}
\usepackage[draft]{hyperref}
\usepackage{color}
\usepackage[]{algorithm2e}
\usepackage{booktabs}
\usepackage[T1]{fontenc}
\usepackage{amsmath}

\usepackage[utf8]{inputenc}
\usepackage[english]{babel}

\usepackage{comment}
\usepackage{textpos}

\IEEEpubid{\makebox[\columnwidth]{978-1-5090-5963-8/17/\$31.00~\copyright~2017
		IEEE \hfill} \hspace{\columnsep}\makebox[\columnwidth]{}}

\begin{document}

\title{A Novel Device-to-Device Discovery Scheme for Underlay Cellular Networks}
\author{\IEEEauthorblockN{Mansour Naslcheraghi${{}^{1}}$, 
		Leila Marandi${{}^{1}}$, Seyed Ali Ghorashi${{}^{1,2}}$, \textit{Senior Member, IEEE}\\}
	\IEEEauthorblockA{1. Department of Electrical Eng, Shahid Beheshti University G. C., Tehran, Iran\\
		2. Cyberspace Research Institute, Shahid Beheshti University G. C., Tehran, Iran\\
		\url{{m.naslcheraghi, l.marandi}@mail.sbu.ac.ir}, 
		\url{a_ghorashi@sbu.ac.ir} }}

\maketitle
\begin{abstract}
Tremendous growing demand for high data rate services such as video, gaming and social networking in wireless cellular systems, attracted researchers' attention to focus on developing proximity services. In this regard, device-to-device (D2D) communications as a promising technology for future cellular systems, plays crucial rule. The key factor in D2D communication is providing efficient peer discovery mechanisms in ultra dense networks. In this paper, we propose a centralized D2D discovery scheme by employing a signaling algorithm to exchange D2D discovery messages between network entities. In this system, potential D2D pairs share uplink cellular users' resources with collision detection, to initiate a D2D links. Stochastic geometry is used to analyze system performance in terms of success probability of the transmitted signal and minimum required time slots for  the proposed discovery scheme. Extensive simulations are used to evaluate the proposed system performance.
\end{abstract}
%% ===================================================================
\IEEEpeerreviewmaketitle 
%{3cm}(4.1cm,-10.5cm)
\begin{textblock*}{3cm}(4.1cm,-12cm)
	\makebox[0.1\columnwidth]{$25^{\rm th}$ Iranian Conference on Electrical Engineering (ICEE2017)}
	
\end{textblock*}
%% ===================================================================
\IEEEpeerreviewmaketitle

\section{Introduction}
Recently, Device-to-Device (D2D) communication as a promising technology of 5G cellular networks, has been emerged to increase the spectral efficiency by reusing the same radio resources among multiple links \cite{D2DLTE}. It is considered as a solution to implement proximity services among multiple devices, such as mobile social networks, public safety and location-based advertisement \cite{ref2}, or for video content delivery \cite{MyIET}. D2D peer discovery is an important procedure to find potential D2D candidates to communicate with each other. Several studies have focussed on different aspects of D2D discovery in recent years. particularly, Long Term Evolution - Advanced (LTE-A) infrastructure is widely used in which, there is no need for extra designs and modifications in current cellular networks. In such systems, peer discovery procedure is under fully control of a central base-station (eNodeB) and is called centralized network-assisted D2D peer discovery. Interference and power control schemes for D2D discovery in Inband cellular by considering user densities, addressed in LTE-A enhancements \cite{ref3}.

In current LTE-A networks, energy consumption aspects of D2D discovery mechanisms which are discussed in the third generation partnership project (3GPP) are studied in \cite{ref4}. A probabilistic model and random access procedure in the LTE-A system that discovers pairs of user equipments (UEs) in a centralized manner have been also proposed in \cite{contrast2} and \cite{ref6}, respectively.  Furthermore, a centrally controlled device discovery scheme tailored to the LTE system is proposed in \cite{ref7}. This scheme \cite{ref7} consists of a comprehensive application layer procedure that enables the device discovery services, and a set of Media Access Control (MAC) layer enhancements that provides a resource request/allocation procedure for discovery transmissions. By utilizing sounding reference signal (SRS) channel, which can be accessed by UEs that are LTE-compliant, a neighbor discovery with D2D channel estimation is proposed in  \cite{ref8}. One approach to allocate radio resources to D2D peer discovery procedures is using LTE-A uplink radio resources. In this regard, physical uplink control channel (PUCCH) of cellular network is used to establish peer discovery signaling messages between BS and potentially D2D pairs. However it can cause inter carrier interference (ICI) and also transmitted signaling messages need to be strong enough to make sure that the BS and D2D candidates can receive the messages, correctly. Boosting transmit power of signaling messages symbols can increase the average transmit power of discovery signal, while maintaining low ICI and cellular PUCCH reception performance \cite{ref9}. Users also may use broadcasting to advertise their presence and service, to discover other devices, autonomously and continuously% \cite{ref13,ref14}. 
[11, 12] 

In this paper, we propose an uplink underlay network-assisted novel scheme for D2D discovery. A network-controlled signaling algorithm is proposed to exchange discovery messages a) between user devices in potential D2D pairs, and b) between D2D pairs and BS. In this scheme, proximity users feedback their identity and channel information to BS in order to provide accurate estimation of the link quality between D2D pairs candidates. In contrast with the existing works in the literature \cite{contrast1,contrast2} in which the transmitted discovery messages assumed to be always successful in delivering to the respective receivers without considering cellular users' influence, we propose a realistic network model based on the Poisson point process (PPP) to include channel state information (CSI) in D2D discovery process by considering imposed interference from cellular users on D2D pairs. We employ stochastic geometry to analyze the system performance in terms of the cumulative distribution function (CDF) of the experienced signal-to-interference ratio (SIR) at D2D receivers and the expected number of needed time slots for D2D discovery processes in a multi-node scenario. 

The rest of paper is organized as follows. In section II, we first delineate our system model and discovery mechanism. Then, we present PPP analysis in section III. The analytic and simulation results are presented in section IV and section V concludes this paper. 

% ---------------------------------
\section{SYSTEM MODEL And D2D DISCOVERY ALGORITHM}
\subsection{System Model}
We consider an infinite cellular network as shown in Fig. 1 with randomly distributed user devices and BSs within the network area. We define ${\Phi _{b}}$ as a Poisson point process (PPP) with density of ${\lambda _{b}}$, which determines the locations of the BSs within the network area. We also define ${\Psi _u}$ as another PPP with density of ${\lambda _{u}}$, which determines the locations of the UEs. Potential D2D pairs are sharing cellular users' resources underlay uplink cellular infrastructure, hence, the mutual interference at D2D receivers is taken into account. The broadband connection is provided for the BSs by central scheduler via wired links. For the notational simplicity and mathematical derivations, we focus on a typical random cell, termed representative cell, and analyze the performance of the proposed D2D discovery algorithm in terms of the transmitted signal's success probability and the expected number of required time slots, to satisfy the system minimum constraints. We neglect co-channel interference and neighboring cell users' influence, for the sake of simplicity, however, the analysis can be applied for multi-cell scenarios, as well. All D2D communications are under fully control of the BS and they transmit signaling messages at the beginning of each time slot. The transmission scheme is time division multiple access (TDMA) which implies, given a time for D2D discovery mechanisms, the whole time is divided into small portions called "time slots" and each user transmits its message at the beginning of each time slot, according to a transmission probability. For a given time slot, one D2D discovery message is allowed to pass at the beginning of this time slot, hence, if two users send their discovery message at the same time, collision occurs and therefore failed messages need to be retransmitted in the next time slots. We further assume that the system operates under interference-limited regime, therefore, the background noise is negligible in comparison with the experienced interference at the receivers.  

\begin{figure}[ht]
	\centering
	\includegraphics[width=0.38 \textwidth]{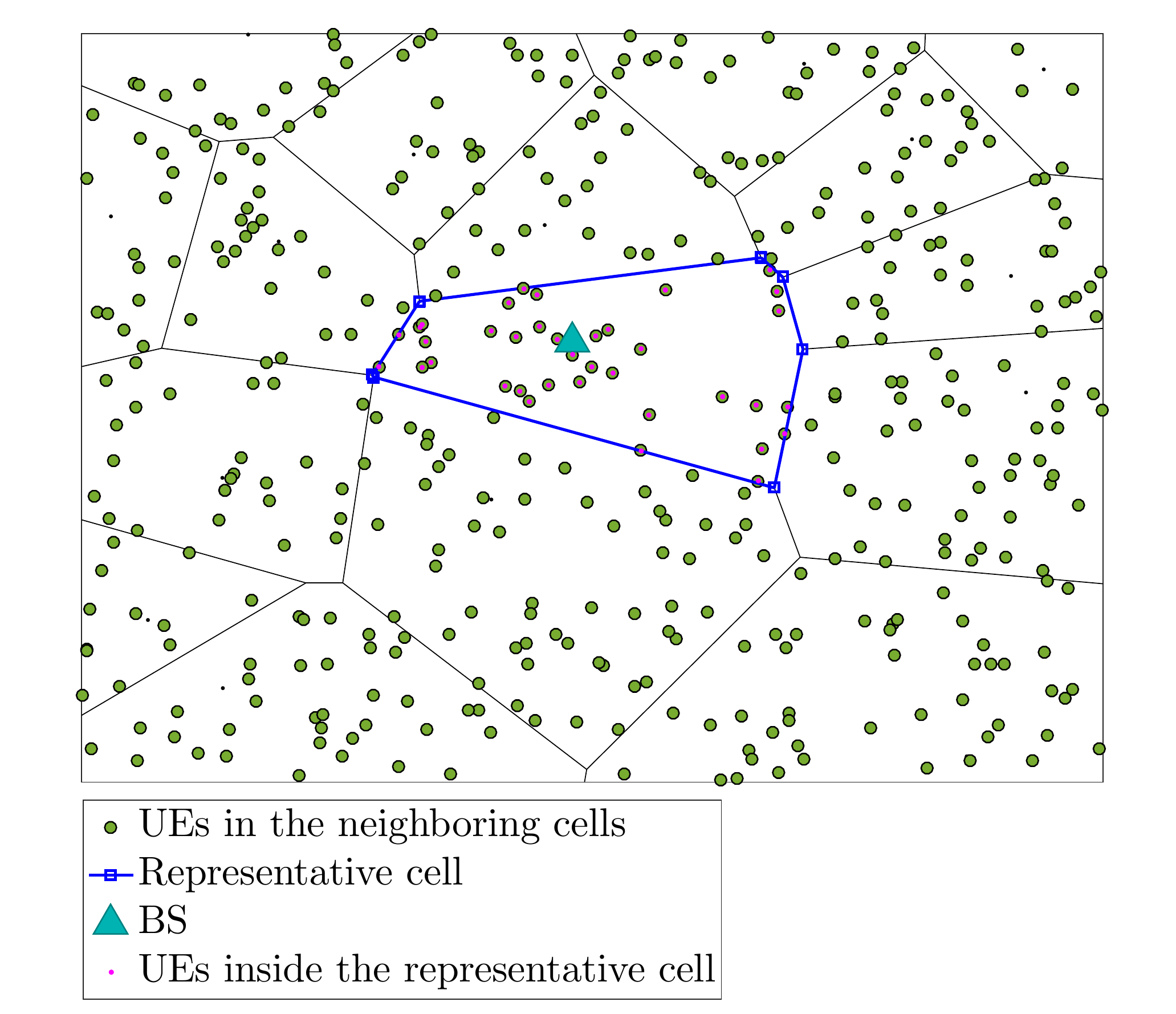}
	\caption{System model with randomly distributed UEs and BSs.}
	\label{overhead}
\end{figure}

\subsection{D2D Discovery Algorithm}
In this subsection, we propose new D2D discovery algorithm by considering the BS as the central management system. Technically, to initiate a D2D link between two users, some signaling messages need to be exchanged, through the network entities. %Even in cellular communications, signaling messages are passing to authorize the user devices to assign necessary resource blocks or or services, e.g. synchronize signaling messages in cellular network. 
In our proposed scheme, since the BS is the central management system to exchange required signaling messages for D2D discovery scheme, the proposed scheme is centralized. We focus on a potential D2D pair ($u_x$, $u_y$) as a representative for all potential D2D pairs and describe the proposed D2D discovery mechanism by considering $u_x$ as transmitter and $u_y$ as its receiver. Assuming that the users can detect other users in their proximity (such a detection can be done by FlashLinQ mechanism proposed by LTE-A for proximity services \cite{ref15}), the proposed algorithm can be defined in five signaling steps as follows.

\begin{enumerate}
	\item User $u_x$ sends a request message to the BS to initiate a D2D link with user $u_y$.

	\item After receiving $u_x$'s transmitted signal at BS, the BS forwards discovery message to the user $u_y$ and schedules the user $u_y$ to listen discovery messages transmitted by the user $u_x$ in its vicinity, and sends acknowledgment signal to user $u_x$. 
	
	\item User $u_x$ sends discovery message to user $u_y$.
	\item After receiving discovery message at user $u_y$, it measures the experienced signal-to-interference ratio (SIR) and sends it back to the BS. 
	\item BS receives measured SIR and queues the D2D pair to initiate D2D communication; if the received SIR satisfy the system threshold, BS initiate users $u_x$ and $u_y$ to initiate D2D communication; otherwise discovery process will fail and the D2D pair need to retransmit their discovery message in the next time slots.
\end{enumerate}

In all the aforementioned steps, when users send the signaling messages to the BS, the received signal at BS should satisfy the system SIR threshold, likewise.
 
\section{Analysis}
In this section we provide the analysis for the proposed scheme by employing stochastic geometry to analyze the transmitted signals' success probability. For the proposed system in section II, there are two key conditions to successfully deliver a discovery message from the transmitter to receiver.

a) Collision: when two users transmit their discovery message at the beginning of a given time slot at the same time, collision occurs and both users have to retransmit their discovery messages. Denoting $N$ as the number of potential D2D pairs, the signal from user $u_x$ will successfully be delivered to its respective receiver, if all other pairs are in idle mode. Each user transmit its discovery signal with a specific transmission probability. Hence we define $P_{nc}$ as the probability of successful signal transmit in a given time slot as
\begin{equation}
{P_{nc}} = {T_o}{(1 - {T_o})^{N - 1}},
\end{equation}
where, $T_o$ is the transmission probability. Now, we aim to maximize $P_{nc}$ by seeking optimal transmission probability:
\begin{equation}
\frac{{\partial {P_{nc}}}}{{\partial {T_o}}} = 0 \Rightarrow {T^*_o} = \frac{1}{N},
\end{equation}
where ${T^*_o}$ is the optimal transmission probability. Hence, denoting $P_{nc}^ * $ as the optimal probability of transmitting a signal without collision, we have:
\begin{equation}
P_{nc}^ *  = \frac{1}{N}{(1 - \frac{1}{N})^{N - 1}}.
\end{equation}

b) SIR: for a given transmitter and its respective receiver, the received signal strength and consequently the SIR level should meet the system design thresholds. This means that a transmitted signal from user $u_x$ can be successfully delivered to receiver $u_y$, if the experienced SIR at receiver $u_y$ is equal or greater than a threshold $\tau$, i.e., 
\begin{equation}
P(SI{R_{xy}} \ge \tau ).
\end{equation}
Eq. (4) also delineates the cumulative distribution function (CDF) of experienced SIR at receiver $u_y$. Since all potential D2D pairs transmit their discovery message at the beginning of a given time slot, the parameter $N$ can be considered as the number of messages which are simultaneously transmitted by potential D2D users. Now, we define success probability for a transmitted signal from transmitter $u_x$ to its respective receiver $u_y$, by considering joint collision and SIR satisfaction. This means that a transmitted signal from user $u_x$ will successfully be delivered to its respective receiver if a) there is no collision in transmitting the signal and b) the received signal at receiver $u_y$ satisfies the system SIR threshold $\tau$, i.e., 
\begin{equation}
{P_{success}} = P(No{\rm{ }}Collision,SI{R_{xy}} \ge \tau ), 
\end{equation}
since the collision process and SIR satisfaction are independent at each D2D pair, we can rewrite the eq. (5) as 
\begin{equation}
{P_{success}} = P_{nc}^ * .P(SI{R_{xy}} \ge \tau ).
\end{equation}
Now we aim to derive the closed form of the CDF for the experienced SIR at the receiver of interest, $u_y$, by employing stochastic geometry approach. The experienced SIR at the receiver $u_y$ due to transmitted signal from user $u_x$ is
\begin{equation}
SI{R_{xy}} = \frac{{{P_t}{h_{xy}}l\left( {x,y} \right)}}{{\sum\limits_{z \in \Phi \backslash \{ x\} } {{P_t}{h_{zy}}l\left( {z,y} \right)} }},
\end{equation}
where, ${\rm{  }}l\left( {x,y} \right) = {\left\| {x - y} \right\|^{ - \alpha }}$ is the standard path loss with exponent of $\alpha$ and $\left\| {x - y} \right\|$ denotes the Euclidean distance between transmitter $u_x$ and its respective receiver, $u_y$. ${\sum\nolimits_{z \in \Phi \backslash \{ x\} } {{P_t}{h_{zy}}{l(z,y)}}}$ is total interference due to the transmitting nodes in set $\Phi$, which is the set of concurrent transmitting nodes. Backslash in eq. (7) implies that the node $u_x$ is excluded from transmitters set. ${{h_{xy}}}$ and ${{h_{zy}}}$ are the fading power coefficients with exponential distribution of mean one, corresponding to the channel gain between transmitter $u_x$ and its respective receiver $u_y$, and the interferer $u_z$, respectively. We consider receiver $u_y$ at origin and transmitting node $u_x$  with a fixed distance of $R$ from $u_y$, which is the nearest user to $u_y$. Hence, all the interferer are in the outside of the circle of radius $R$. Now, by denoting $I$ as the total interference, i.e., $I = \sum\nolimits_{z \in \Phi \backslash \{ x\} } {{h_{zy}}l(z,y)} $ , and using Campbell’s theorem \cite{Campell},  the Laplace transform of the interference \cite{stochastic} can be defined as: 
\begin{align}
%\begin{array}{l}
{\mathscr{L}_{I}\left( s \right)} &= \mathscr{L} \left( {\sum\limits_{z \in \Phi } {{h_{zy}}l\left( {z,y} \right)} } \right)\\\notag
&=E\left( {\prod\limits_{z \in \Phi } {{e^{ - s{h_{zy}}l\left( {z,y} \right)}}} } \right)\\\notag
&=\exp \left( { - \pi \lambda_{u} \Gamma \left( {1 + \delta } \right)\Gamma \left( {1 - \delta } \right){s^\delta }} \right)\\\notag
&\buildrel \Delta \over = \exp \left( { - \lambda_{u} G\left( {s,\alpha } \right)} \right), 
%\end{array}
\end{align}
where, $G\left( {s,\alpha } \right) = \frac{{{\pi ^2}\delta {s^\delta }}}{{\sin \left( {\pi \delta } \right)}}$, $\delta  \buildrel \Delta \over = \frac{2}{\alpha }$ and $\Gamma (.)$ is the gamma function. Now, the final expression for the CDF of SIR at receiver $u_y$ can be defined as 
\begin{align}
P\left( {SI{R_{xy}} \ge \tau } \right)& = {L_I}\left( {\tau {R^\alpha }} \right)\\&\notag = \exp \left( { - \lambda_u G\left( {\tau {R^\alpha },\alpha } \right)} \right).
\end{align}
By substituting equations (3) and (9) in eq. (6), the final expression for success probability can be derived. Now, we define random variable $X$ denoting the number of successful transmissions in given $n$ time slots. The probability that there is at least one successful discovery signal is
\begin{equation}
p\left( {X \ge 1} \right) = 1 - {\left( {1 - {p_{success}}} \right)^n}.
\end{equation}
Now, we define another system design parameter $\eta$, as the minimum required success probability for requiring at least one time slot for a successful discovery message. We have: 
\begin{equation}
p\left( {X \ge 1} \right) \ge \eta.  
\end{equation}
By substituting equations (3) and (9) in (6) and then eq. (6) in (10) and then (10) in (11), respectively, and some simple manipulations, the maximum number of required time slots for the proposed D2D discovery mechanism can be defined as
\begin{align}
n \le \frac{{\ln (1 - \eta )}}{{\ln (1 - \frac{1}{N}{{(1 - \frac{1}{N})}^{N - 1}}exp( - {\lambda _u}G(\tau {R^\alpha },\alpha )))}},
\end{align}
where, 
\begin{align}
G\left( {\tau {R^\alpha },\alpha } \right) = \frac{{{\pi ^2}\delta {{\left( {\tau {R^\alpha }} \right)}^\alpha }}}{{\sin \left( {\pi \delta } \right)}};{\rm{ }}\delta  \buildrel \Delta \over = \frac{2}{\alpha };{\rm{ }}\alpha> 2.\notag
\end{align}
$\alpha> 2$ is the requirement of the stochastic geometry approach addressed in \cite{stochastic}. 

%sample equations: 
%\begin{equation}
%T_{{u_i}}^{HD} = \sum\limits_{{u_j} \in {\rm A}} {WE\left[ {1 + {{\log }_2}(1 + SIN{R_j})} \right]}, 
%\end{equation}
%
%\begin{align}
%T_{{u_i}}^{FD} =& WE\left[ {1 + {{\log }_2}(1 + SIN{R_i})} \right] \notag\\&+ \sum\limits_{{u_j} \in B} W{E\left[ {1 + {{\log }_2}(1 + SIN{R_j})} \right]}, 
%\end{align}

\section{Analytic and Simulation Results} 
In this section, we provide Monte-Carlo simulations to evaluate system performance. We focus on an isolated single cell of the voronoi tessellation network (Fig. 1) and neglect inter-cell interference. We assume standard microcell channel model by considering log-normal shadow fading effect. The rest of simulation parameters are summarized in Table 1. In the wireless D2D network, described in section II, Users initiate for D2D communication with the nearest neighbor in the vicinity. We choose potential D2D pairs, according to given distance threshold and label them by IDs as $\{ D{D_1},D{D_2},...,D{D_N}\}$. The key parameter in this system is collision detection in simultaneous transmissions by D2D pairs in the beginning of each time slot. As we analyzed in section III, the optimal transmission probability in the beginning of each time slot is given by ${T_o} = \frac{1}{N}$ (eq. 2). Since the channel access model is TDMA, to simulate collision detection, at the beginning of each time slot, each user throws its dice to pick up a number between [1 $N$], if the chosen number is equal to the respective D2D pair's ID, then, the D2D transmitter will be authorized to transmit its discovery signal to its respective receiver. And if the received signal in the receiver can satisfies system thresholds ($\tau$), D2D pair succeed in the collision process and will proceed to the next step of the discovery algorithm. 
\begin{table}[h]
	\caption {Simulation parameters} \label{tab:title}
	\begin{footnotesize}
		\small{
			\centering
			\begin{tabular}{l|l}
				\toprule
				Parameter & Values\\
				\midrule
				UEs density ($\lambda_u$) & [$10^{-3}$ $10^0$]\\
				BSs density ($\lambda_b$) & 0.2\\
				Under consideration D2D pairs ($N$) &  2, 4, 6, 8\\
				Required success probability ($\eta$)  &  0.9\\
				DUEs SINR threshold ($\tau$) & [-20 20] dBm\\
				Path loss exponent ($\alpha$) & 4\\
				Distance between DUEs($R$) & 30 m\\
				User transmit power ($P_t$) & 23 dBm\\
				BS transmit power ($P_{bs}$) & 40 dBm\\
				Log-normal shadow fading & 4 dB standard deviation\\
				Monte-Carlo iterations & 1000\\
				\bottomrule
			\end{tabular}
			\label{tbl:params1}}
	\end{footnotesize}
\end{table}

In what follows, we first describe the analytic results and possible tradeoffs in Fig. 1, 2, and 3 and then describe the simulation results for delineated system model in section II. 

Fig. 2 demonstrates the impact of $\lambda_u$ on the CDF of SIR for different system SIR thresholds ($\tau$). As can be seen in this figure, by increasing the density of users ($\lambda_u$) within the cell area, which corresponds to more potential D2D pairs within the cell area, the probability of successful discovery messages transmissions with respect to system SIR threshold, decreases, because, the experienced interference at D2D receivers increases. In other hand, increasing system SIR thresholds which corresponds to guaranteeing higher D2D link quality, leads to decreasing the probability of successful discovery messages deliveries. Therefore, there is a trade off between D2D link quality and successful discovery messages deliveries. i.e., guaranteeing higher D2D link quality, the lower is the probability of successful delivery of discovery messages and guaranteeing the lower D2D link quality, the higher is the successful delivery of discovery message. Fig. 3 shows the impact of system SIR threshold on the CDF of SIR. Similar explanations for Fig. 2 are valid for Fig. 3. 

\begin{figure}[b]
	\centering
	\includegraphics[width=0.43 \textwidth]{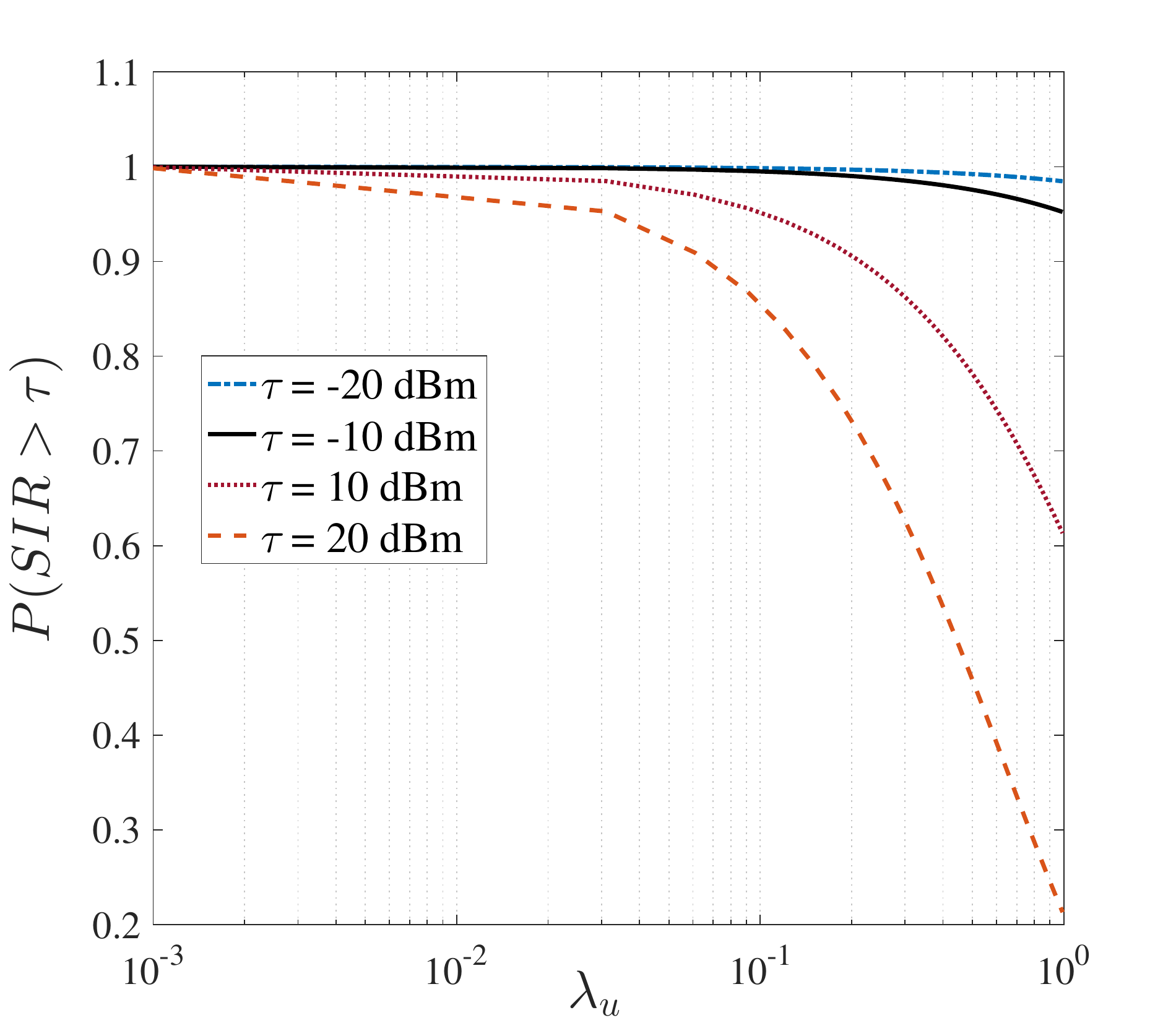}
	\caption{CDF of SIR versus network density $\lambda_u$.}
\end{figure}

Fig. 4 demonstrates the main success probability defined in eq. (6). In this figure, we consider limited number of D2D pairs ($N$) with the impact of joint collision detection and SIR satisfaction. As can be seen from this figure, collision detection has a major impact on the success probability, due to simultaneous transmission at the beginning of time slots. By increasing the number of D2D pairs in a given time slot, to transmit their desired discovery message, the impact of collision detection is more accented. 

Fig. 5 shows the CDF of the minimum required number of time slots for successful D2D discovery process. As expected, by increasing the number of potential D2D pairs in a specific $\lambda$ and $\tau$ , the required time slots increases due to more collisions in channel access. In other hand, by increasing the density of the network, since D2D pairs are sharing uplink resources of the cellular users, there is high interference on D2D receivers, which leads to more fails in D2D discovery messages deliveries. Fig. 6 demonstrates the number of required time slots for different number of D2D pairs as derived in eq. (12). Similar explanations for Fig. 5 are also valid for Fig. 6.  

\begin{figure}[ht]
	\centering
	\includegraphics[width=0.43 \textwidth]{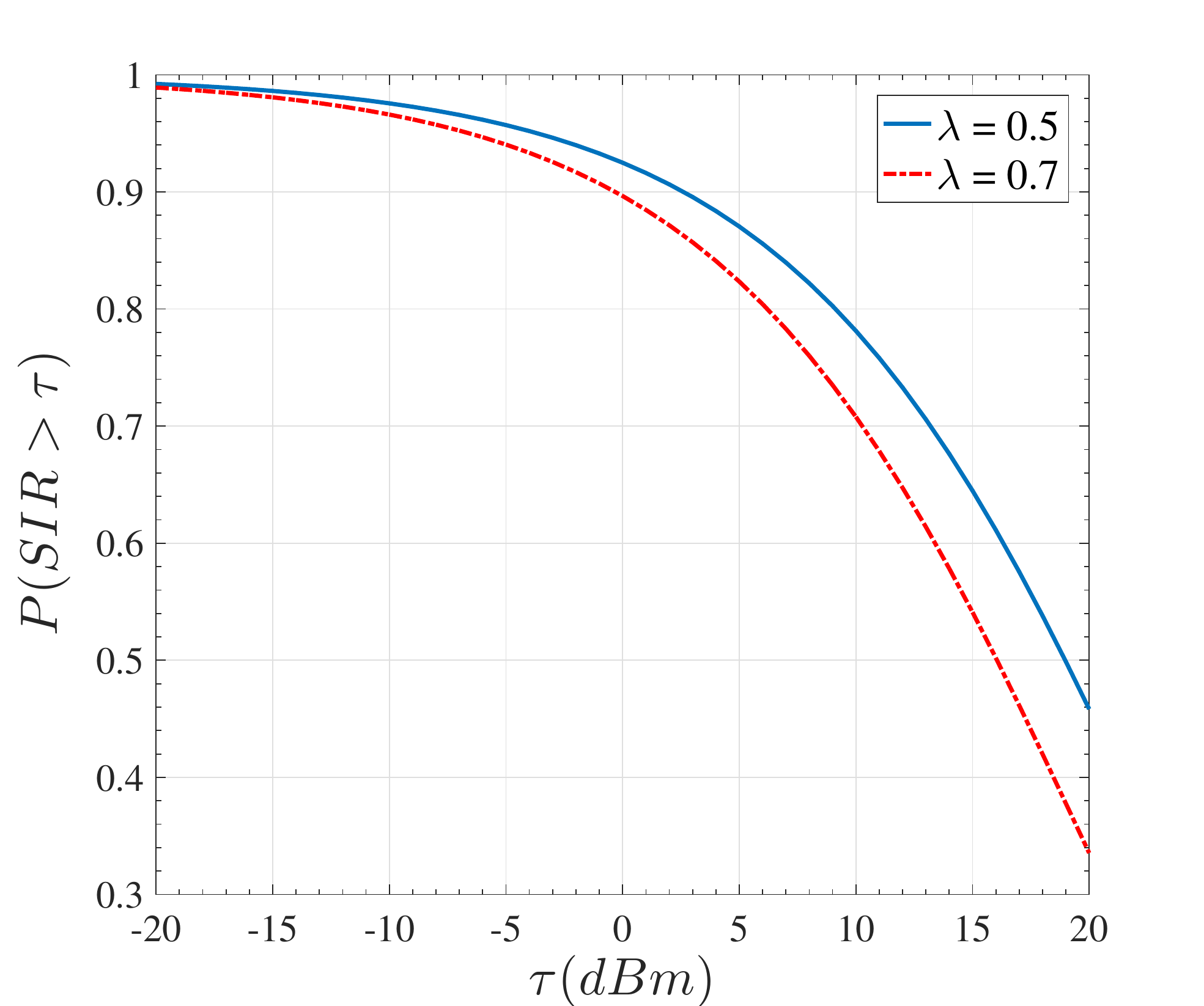}
	\caption{CDF of SIR versus $\tau$ for $\lambda_u=0.5, 0.7$.}
\end{figure}
\begin{figure}[h]
	\centering
	\includegraphics[width=0.41 \textwidth]{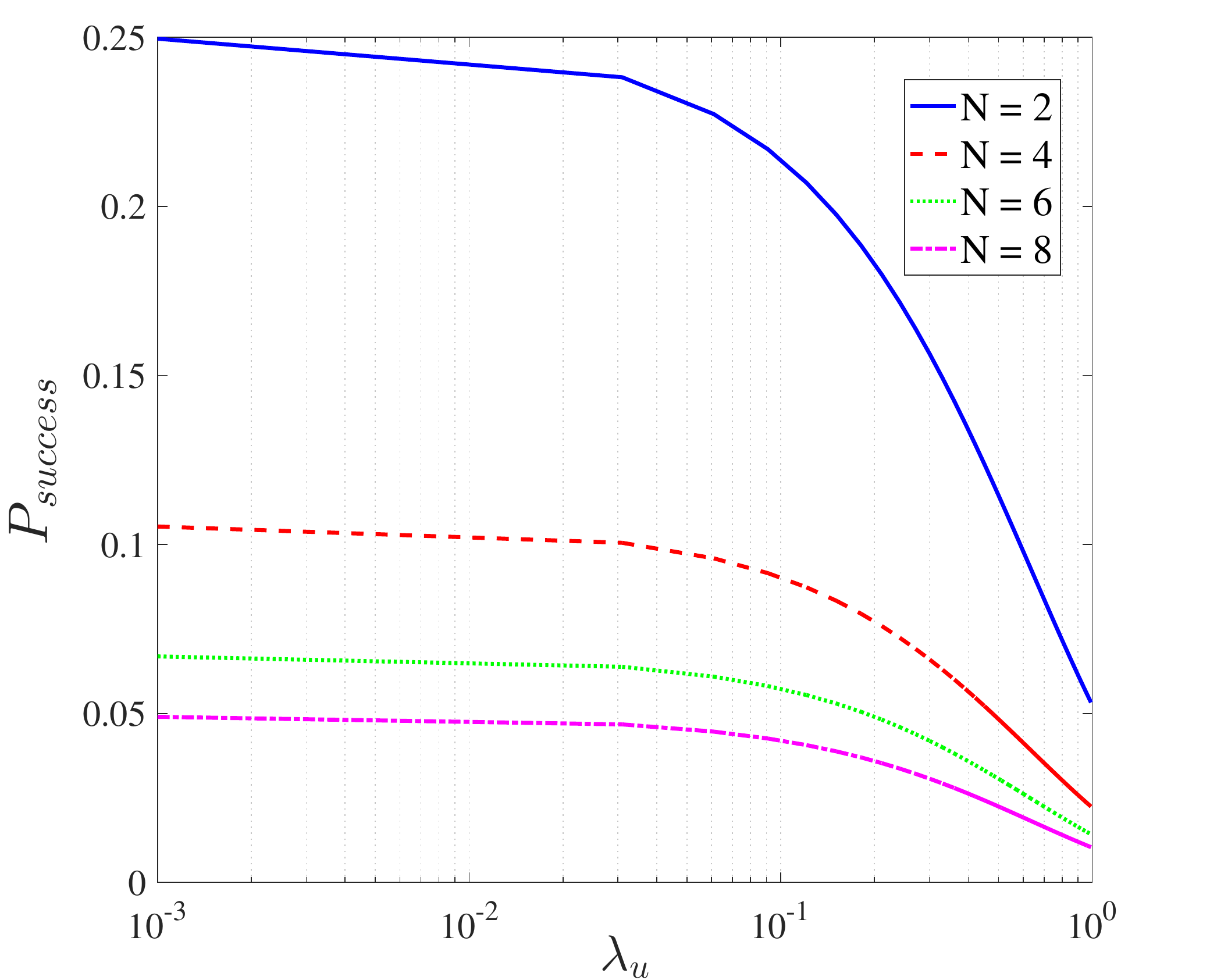}
	\caption{$P_{success}$ versus $\lambda_u$ for $\tau=20$ dBm.}
\end{figure}
          
\begin{figure}[h]
	\centering
	\includegraphics[width=0.43 \textwidth]{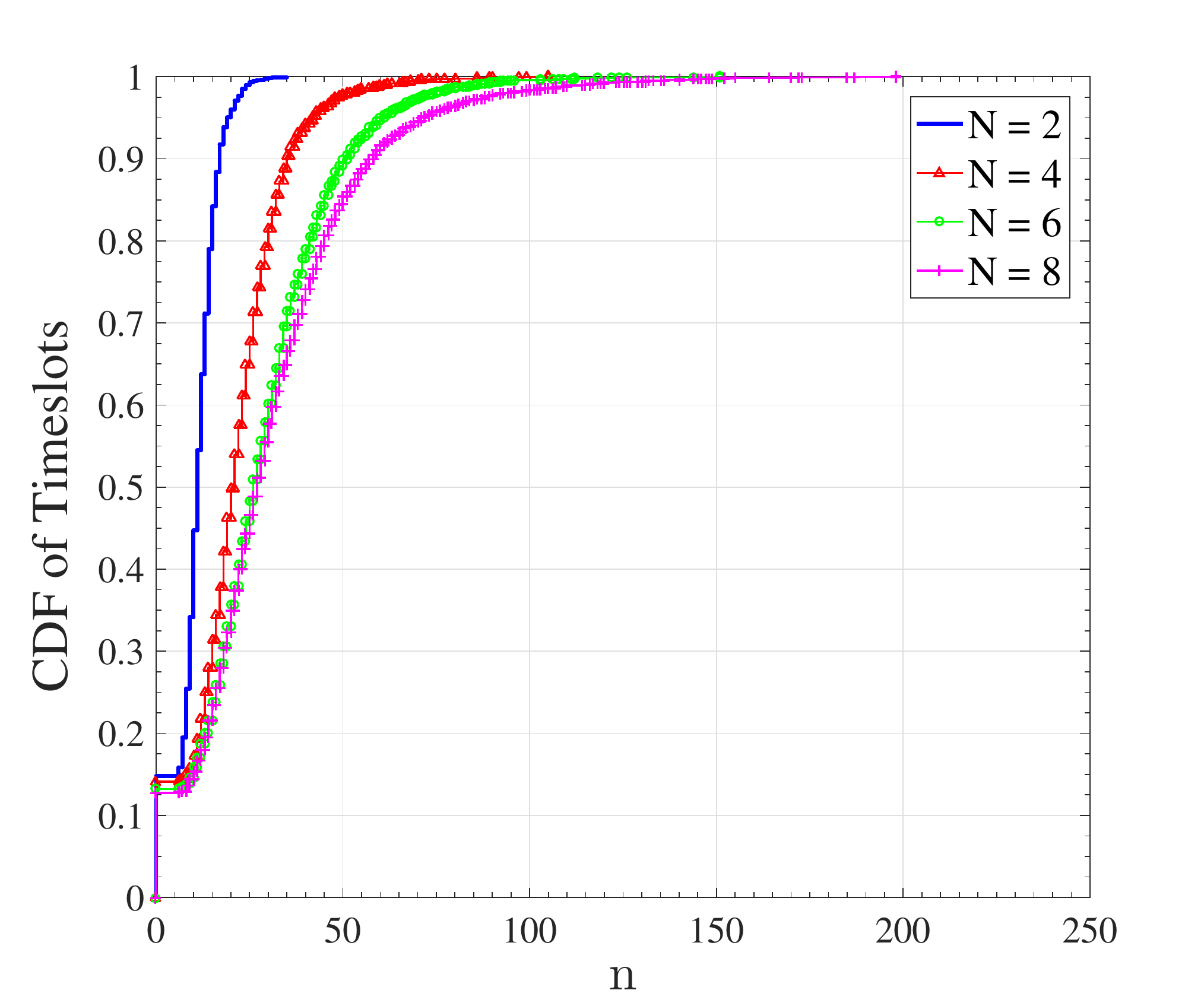}
	\caption{CDF of time slots for $\lambda_u = 0.4 $ and $\tau=20$ dBm.}
\end{figure}

\begin{figure}[h]
	\centering
	\includegraphics[width=0.43 \textwidth]{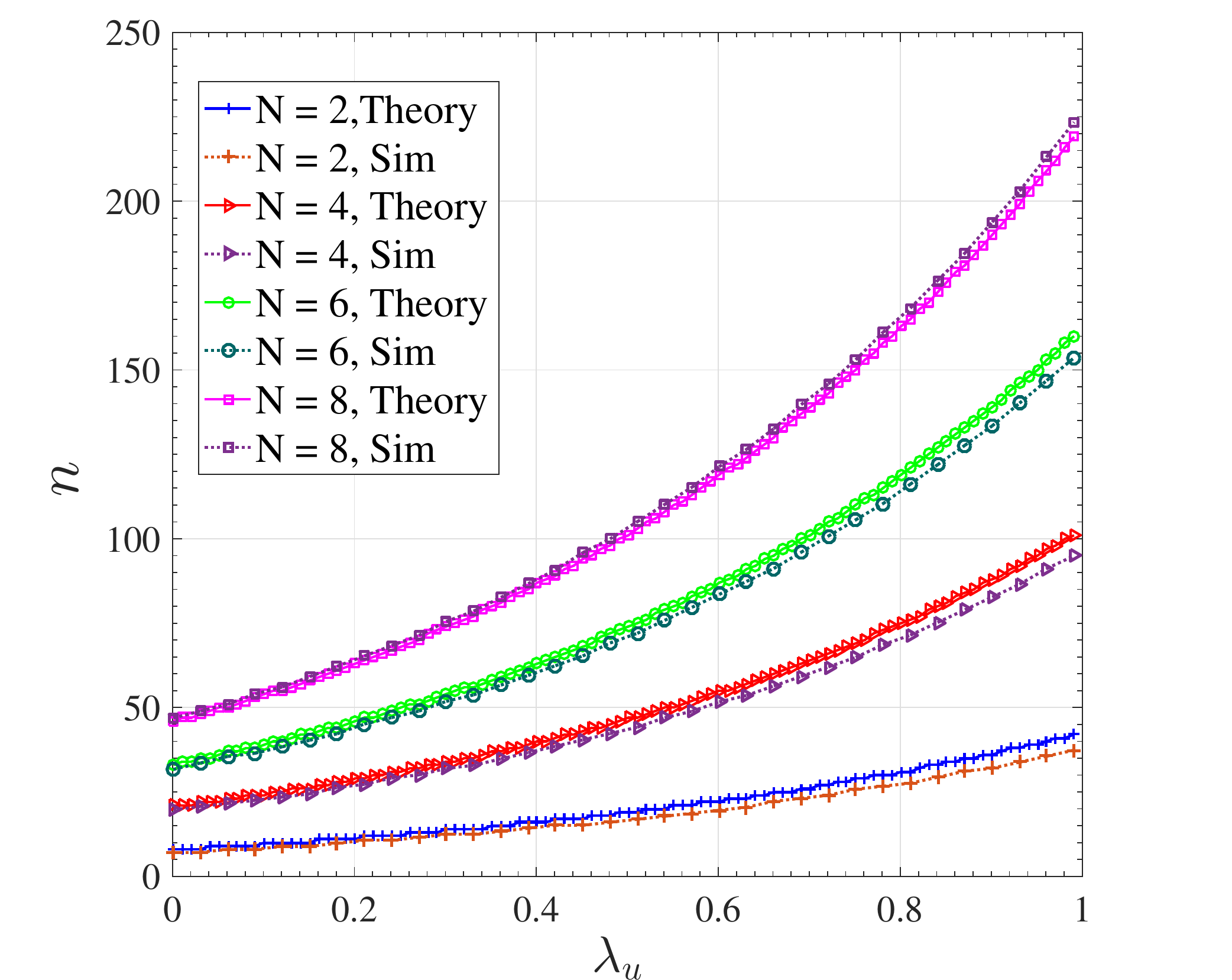}
	\caption{Required time slots versus $\lambda_u$ for $\tau=20$ dBm.}
\end{figure}

\section{CONCLUSIONS}
In this paper, we proposed a novel scheme for D2D discovery by employing a centralized signaling algorithm for exchanging discovery messages between network entities. We used stochastic geometry to analyze and evaluate the realistic performance of the proposed scheme. For this system, proposing new efficient collision avoidance algorithms in TDMA channel access can promises even high reliability to deliver D2D discovery messages.   

\bibliographystyle{IEEEtran}

\begin{thebibliography}{9}

	\bibitem{D2DLTE} 
	K. Doppler, M. Rinne, C. Wijting, C. B. Ribeiro, and K. Hugl, "Device-to-device communication as an underlay to LTE-advanced networks," \textit{IEEE Communications Magazine}, vol. 47, pp. 42-49, 2009.
	
	\bibitem{ref2}
	3GPP, "TR 22.803, Feasibility study for Proximity Services (ProSe)," ed, June 2013.
	
	\bibitem{MyIET}M. Naslcheraghi, SA. Ghorashi, and M. Shikh-Bahaei, "Full-Duplex Device-to-Device Communication for Wireless Video Distribution," \textit{IET Communications}, 2017.
	
	
	\bibitem{ref3}
	D. Li and Y. Liu, "Performance analysis for LTE-A device-to-device discovery," \textit{IEEE PIMRC}, pp. 1531-1535, 2015.
	
	\bibitem{ref4}
	A. Prasad, A. Kunz, G. Velev, K. Samdanis, and J. Song, "Energy-efficient D2D discovery for proximity services in 3GPP LTE-advanced networks: ProSe discovery mechanisms," IEEE vehicular technology magazine, vol. 9, pp. 40-50, 2014.
	
	\bibitem{contrast1}
	P. Nguyen, P. Wijesinghe, R. Palipana, K. Lin, and D. Vasic, "Network-assisted device discovery for LTE-based D2D communication systems," \textit{IEEE ICC}, pp. 3160-3165, 2014.
	
	\bibitem{contrast2}
	Thanos, Anastasios, Serveh Shalmashi, and Guowang Miao. "Network-assisted discovery for device-to-device communications," \textit{IEEE Globecom Workshops}, 2013.
	
	\bibitem{ref6}
	K. W. Choi and Z. Han, "Device-to-device discovery for proximity-based service in LTE-advanced system," \textit{IEEE Journal on Selected Areas in Communications}, vol. 33, pp. 55-66, 2015.
	
	\bibitem{ref7}
	D. Tsolkas, N. Passas, L. Merakos, and A. Salkintzis, "A device discovery scheme for proximity services in LTE networks," \textit{IEEE ISCC}, pp. 1-6, 2014.
	
	\bibitem{ref8}
	H. Tang, Z. Ding, and B. C. Levy, "Enabling D2D communications through neighbor discovery in LTE cellular networks," \textit{IEEE Trans. on Signal Processing}, vol. 62, pp. 5157-5170, 2014.
	
	\bibitem{ref9}
	J. H. Song, D. H. Lee, W. J. Hwang, and H. J. Choi, "A selective transmission power boosting method for D2D discovery in 3GPP LTE cellular system," \textit{IEEE ICTC}, pp. 267-268, 2014.
	
%	\bibitem{ref10}
%	S. B. Seo, J. Y. Kim, and W. S. Jeon, "Robust and fast device discovery in OFDMA-based cellular networks for disaster environment," 18th IEEE International Conference on Advanced Communication Technology (ICACT), pp.498-502, 2016.
%%	
%	\bibitem{ref11}
%	W. Hwang, D. Lee, and H. J. Choi, "A new channel structure and power control strategy for D2D discovery in LTE cellular network," 20th IEEE Asia-Pacific Conference on Communication (APCC), pp. 150-155, 2014.
	
%	\bibitem{ref12}
%	A. Prasad, K. Samdanis, A. Kunz, and J. Song, "Energy efficient device discovery for social cloud applications in 3GPP LTE-advanced networks," IEEE Symposium on Computers and Communications (ISCC), pp. 1-6, 2014.
%	
	\bibitem{ref13}
	K. W. Choi, H. Lee, and S. C. Chang,"Discovering Mobile Applications in Device-to-Device Communications: Hash Function-Based Approach," \textit{IEEE VTC Spring}, pp. 1-5, 2014.
	
	\bibitem{ref14}
	S. Jung and S. Chang, "A discovery scheme for device-to-device communications in synchronous distributed networks," \textit{IEEE International Conference on Advanced Communication Technology}, pp. 815-819, 2014.
	
	\bibitem{ref15}
	X. Wu, S. Tavildar, S. Shakkottai, T. Richardson, J. Li, R. Laroia and A. Jovicic, "FlashLinQ: A synchronous distributed scheduler for peer-to-peer ad hoc networks," \textit{IEEE/ACM Trans. on Networking} (TON), vol. 21, pp.1215-1228, 2013. 
	
	\bibitem{Campell}
	H. ElSawy; A. Sultan-Salem; M. S. Alouini; M. Z. Win, "Modeling and Analysis of Cellular Networks using Stochastic Geometry: A Tutorial," \textit{IEEE Communications Surveys and Tutorials}, vol.PP, no.99, pp.1-1.
	
	\bibitem{stochastic}
	F. Baccelli and B. Blaszczyszyn. "Stochastic geometry and wireless networks, Volume II-Applications," 2009.
	
\end{thebibliography}

\end{document}